\documentclass{article}
\usepackage{amsfonts}
\usepackage{graphicx}
\date{}
\begin{document}
\title{A note on some mathematical models on the effects of Bt-maize exposure: novel results } 
\author { Francesco Camastra, Angelo Ciaramella, Antonino Staiano \\
               Department of Applied Science, \\
              University of Naples Parthenope \\
             Centro Direzionale Isola C4 - 80143 Napoli, Italy\\
               \small{francesco.camastra@uniparthenope.it, angelo.ciaramella@uniparthenope.it,} \\
                      \small{antonino.staiano@uniparthenope.it} 
}
\maketitle
%
%
\begin{abstract}
Some mathematical models for the estimation of the effects of Cry1Ab and Cry1F Bt-maize exposure in the biodiversity are discussed. Novel results about these models are obtained and described in the note. The exact formula for  the proportion of the population that suffers mortality exposed to Cry1Ab pollen, underlining its  dependence  on the margin from the Bt crop edge, is derived.
In addition,  regarding Cry1F pollen effects, it is proposed a procedure, using a probabilistic and statistical approach, that computes the width of the non Bt-stripes used as mitigation measures.
Finally, it has been derived a lower bound, using probabilistic consideration, on the species 
sensitivity of  Lepidoptera.
\end{abstract}
\section{Introduction}
In the last years, the cultivation of genetically modified plants (GMP) is very
widespread in the world, in particular in America and Asia. On the other side, the debate in the scientific community and in the public opinion about the GMP effects is becoming harder and harder
\cite{Mendelsohn03,Banks08,Sanvido11}.
It is our belief, that one of the causes of the debate depends on the lack of available mathematical models that allow to assess quantitatively the effects of the GMP cultivation on the biodiversity.   
To this purpose, recently appeared some approaches that discuss the effects of Bt-maize \cite{Felke11,Meissle11,Szekacs12,Yu11}, and two mathematical models for estimating the effects of Cry1Ab and Cry1F Bt-maize \cite{Perry10, Perry12}, on non-target lepidoptera. To the best of  our knowledge \cite{Lang11, Perry11}, these latter models are the first and the unique mathematical models that assess quantitatively the effects of Bt-crop on the biodiversity. For this reason, we have analyzed in detail the models \cite{Perry10, Perry12}, obtaining novel results. In particular, regarding the Cry1Ab maize \cite{Perry10} has been derived the exact formula, and not the approximated as in \cite{Perry10}, for the proportion of the population that suffers mortality $P$. The exact formula allow to enlight the effective dependence of  $P$ on the margin from the Bt crop edge $D$. 
\\
Regarding Cry1F model \cite{Perry12}, we have derived a procedure, using a probabilistic and statistical approach, that allows to fix automatically the width of the non Bt-stripes used as mitigation measures. Moreover, we have shown that the values of width, proposed  in \cite{Perry12}, does not implement the worst-case method, recommended by the directive of the European Community. Finally, it has been derived a lower bound, using probabilistic consideration, on the species 
sensitivity of the hypothethical Lepidoptera considered in \cite{Perry12}.
\\
The work is organized as follows:
In Section \ref{Cry1ab} the model for the estimation of Cry1Ab Bt-maize  effects is discussed and
the exact formula for   the proportion of the population that suffers mortality $P$ is derived; In Section \ref{Cry1F} the model for the estimation of Cry1F is analyzed and the  procedure that 
computes  the width of the non Bt-stripes is presented; finally, some conclusions are drawn in
Section \ref{concl}.
\section{Estimation of Cry1Ab Bt-maize  effects}\label{Cry1ab}
Perry et al. \cite{Perry10} developed a mathematical model for estimating
the effects of exposure of three non-target Lepidoptera (e.g., \emph{Inachis io}, 
\emph{Vanessa atalanta}, \emph{Plutella xylostella}) to Bt-maize pollen 
expressing the protein Cry1Ab. 
They computed the estimated proportion  of the population of a non-target
Lepidoptera that suffers mortality, hereinafter denoted by $P$ for convenience. $P$   is given by
\begin{equation}\label{eq0}
P= \frac{yzvxa (25 eh \sqrt C + fD \mu)} {(25 e  \sqrt C +fD)},
\end{equation}
where the parameters  $h$ (within-crop mortality),  $x$ (physical effects), $a$ (temporal coincidence), 
$z$ (maize cropping), $v$ (utilization rate), $y$ (host plant in arable), 
$e$ (host plant within-crop),  $f$ (host plant in margin), $C$ (size of maize fields), 
$D$ (width of margin) are experimentally measured or taken by the literature; whereas the
parameter  $\mu$, depending on $D$, has to be properly computed.
\\
Following Perry et al., before computing $\mu$ it is necessary to recall the so-called \emph{margin mortality} $g(E)$ 
where $E$ denotes the distance from the edge of the crop, that is given,  
for  larvae of \emph{Inachis io} and \emph{Vanessa atalanta}, by
\begin{equation}\label{equa1}
g(E) = \frac{\exp(-0.359E)}{33.25 + \exp(-0.359 E)};
\end{equation}
whereas for larvae of \emph{Plutella xylostella}  is given by
\begin{equation}\label{equa2}
g(E) = \frac{\exp(-0.349 E)}{55.33 + \exp(-0.349 E)}.
\end{equation}
That being said, Perry et al. declared  that $\mu$ is obtained by averaging the value of $g(E)$
over the margin. In practice, $\mu$ is obtained by the numerical integration of equations 
(\ref{equa1}) or (\ref{equa2}), between the values $E=0$  and $E=D$. In this way, they
obtained an approximate estimate of $\mu$ and, consequently of the proportion of population that
suffers mortality $P$. Besides, not having an exact formula that states $\mu$ in function of $D$,
they cannot establish the effective dependence of the proportion $P$ on the margin $D$.  
\subsection{The exact proportion  suffering mortality formula}
Now, we show that, differently of what 
claimed by Perry et al., an exact value for $\mu$ and hence $P$ can be derivable. 
The parameter $\mu$ is given by
\begin{equation}\label{equa3}
\mu= \frac{1}{D} \int_{0}^D g(E) dE = \frac{1}{D} \int_{0}^D \frac{\exp(- \gamma E)}{\delta + \exp(- \gamma E)}  dE, 
\end{equation}
where $\gamma$ and $\delta$ are $0.359$ and $33.25$ for the larvae of  
\emph{Inachis io} and \emph{Vanessa atalanta},  whereas are $0.349$ e $55.33$ for the larvae of  \emph{Plutella 
xylostella}.
\\
The integral of the equation (\ref{equa3}) can be solved exactly:
 \begin{eqnarray}
\mu & = & \frac{1}{D} \int_{0}^D   \frac{\exp(- \gamma E)}{\delta + \exp(- \gamma E)} dE \nonumber \\
     & = &  \frac{1}{D} \int_{0}^D [ 1 -  \frac{\delta}{\delta + \exp(-\gamma E)}] dE \nonumber \\  
    & = & \frac{1}{D} \int_{0}^D [ 1 -  \frac{\delta \exp(\gamma E)}{\delta \exp(\gamma E)+1}] dE \nonumber \\  
    & = & \frac{1}{D}\displaystyle[E - \frac{1}{\gamma}\ln (1+\delta \exp(\gamma E))]_{0}^{D} \nonumber \\
     & = &  \frac{1}{D}[ D  - \frac{1}{\gamma}\ln (1+\delta \exp(\gamma D)) +  \frac{1}{\gamma} \ln (1+\delta)] \nonumber \\
     &  = &  \frac{1}{D}[ D  - \frac{1}{\gamma}\ln \frac{1+\delta \exp(\gamma D)}{1+\delta}].
\end{eqnarray}
Plugging last formula in the equation (\ref{eq0}), we obtain the exact formula for computing 
the proportion of population that
suffers mortality $P$
\begin{equation}\label{eqfin}
P= \frac{yzvxa \{25 eh \sqrt C + f [ D  - \frac{1}{\gamma}\ln \frac{1+\delta \exp(\gamma D)}{1+\delta}]\}} {25 e  \sqrt C +fD}.
\end{equation}
The previous equation allows us to predict quantitatively, and not qualitatively as 
performed in Perry et al.'s  work,   the proportion of population that
suffers mortality $P$ at  large distance $D$ from the crop edge.
We pass to study the dependence on $D$ of the proportion $P$. 
When $D=0$, i.e.,  when the non-target Lepidoptera is 
on the crop edge, $P$ becomes:
\begin{equation}\label{p0}
P(0)=yzvxah.
\end{equation}
When $D$ goes to the infinity the equation (\ref{eqfin}) reduces to:
\begin{equation}\label{eqfin2}
P(D) \approx \frac{yzvxa ( 25 eh \sqrt C + \frac{f}{\gamma}  \ln \frac{1+\delta}{\delta} ) } {25 e  \sqrt C +fD} \approx O{\left(\frac{1}{D}\right)}.
\end{equation}
This means that to the infinity $P$ goes to $0$ as $\frac{1}{D}$ that implies
that even for large $D$, $P$ is not negligible.  
\section{Estimation of Cry1F Bt-maize effects}\label{Cry1F}
Perry et al. \cite{Perry12} developed a mathematical model for estimating
the effects of exposure of five hypothetical non-target Lepidoptera species 
to Bt-maize pollen expressing the protein Cry1F. 
In the work the estimated proportion of the population of a non-target Lepidoptera that 
suffers mortality, $P$, is given by the equation (1), that for our convenience, we recall:
\[
P(D)= \frac{yzvxa(25eh\sqrt{C}+fD\mu)}{25e\sqrt{C}+fD}.
\]
The previous equation combines: (a) \textit{small scale parameters}, namely, \emph{e} (host plant within-crop), \emph{f} (host plant in margin), \emph{C} (size of maize field), \emph{D} (width of margin); (b) \textit{large scale parameters}, namely, \emph{y} (host plant in arable), \emph{z} (maize cropping), 
\emph{x} (physical effects), \emph{a} (temporal coincidence); (c) \textit{mortality parameters}, 
namely, \emph{h} (within crop mortality) and $\mu$ (average mortality within a margin of any particular width $D$).
Small and large scale parameters are taken from literature or experimentally measured. The mortality parameters, instead, have to be properly computed, hence in the following the focus will be on the way this is accomplished.
\\
Equation (\ref{eq0}) is derived for a range of five hypothetical non-target Lepidoptera species rather than a specific one. 
To this aim,  Perry et al.  introduce a further \emph{mortality parameter}, 
\emph{m}, representing a range of species sensitivities for the hypothetical non-target Lepidoptera. 
Sensitivity is expressed by the $LC50$ values for maize $1507$, i.e., the lethal 
concentration value that kills on average half of the larvae of the instar considered, 
measured in pollen grains per $cm^{2}$. 
The parameter \emph{m} affects the way \emph{h} and $\mu$ are computed. Let us focus on a step-by-step mortality parameter derivation 
\cite{Perry12}.
The starting point is a mortality-dose laboratory-derived bioassay relationship in 
which a logit-transformed probability of mortality, $P$, is regressed on a logarithmically transformed dose, $d$:
\begin{equation}\label{logitPd}
logit(P) = \alpha + 2.473\;log_{10} d.
\end{equation}
Here is where the parameter \emph{m} comes into play. The intercept $\alpha$ is determined by the sensitivity of the species to the Cry1F protein, for which $logit(P)=0$. Five sensitivity values, corresponding to 
five hypothetical species, are considered in 
\cite{Perry12} 
and denoted as \emph{worst-case, extreme} $(m = 1.265)$, \emph{very high} $(m=14.36)$, \emph{high} $(m=163.2)$, \emph{above-average} $(m= 1853)$, and \emph{below-average} $(m=21057)$. These mortality-dose relationships are then combined with a field-derived regression of logarithmically transformed dose, $d$, on distance $E$, from the nearest source of the pollen:
\begin{equation}\label{log10d}
log_{10}d = 2.346 - 0.145E,
\end{equation} 
to derive a linear mortality-distance relationship for mortality of larvae in the margin, on the logit scale. So doing, for each species sensitivities the mortality-distance relationships are derived, 
from equations (\ref{logitPd}) and (\ref{log10d}), as

\begin{equation}\label{logit}
logit(P) = \beta_0 - 0.3586E,
\end{equation}
where $\beta_0 = \alpha + 5.8017$ and whose values, for each of the considered sensitivities, are shown in 
Table \ref{tab1}.
\begin{table}[h]
\begin{center}
\begin{tabular}{p{3cm}r}
\hline\hline
\\
Sensitivity &   $\beta_0$ \\ [1ex]
\hline 
\textit{extreme}  & $5.5492$ \\ [1ex]
\textit{very high}   & $2.9399$ \\ [1ex]
\textit{high}   & $0.3297$ \\ [1ex]
\textit{above-average} & $-2.2798$ \\ [1ex]
\textit{below-average} & $-4.8901$ \\ [1ex]
\hline
\end{tabular}
\end{center}
\caption{Intercepts, $\beta_0$, for the linear mortality-distance relationships in 
equation (\ref{logit}), corresponding to the considered species sensitivities.}
\label{tab1} 
\end{table}
\\
Taking the inverse of the logit function, we return to the natural scale thus obtaining the estimated probability of mortality $g(E)$, for a larva at distance $E$ into the margin from the nearest source of pollen at the edge of the field:
\begin{equation}\label{P}
g(E) = logit^{-1}(P) = \frac{\exp(\beta_0-0.3586E)}{1+\exp(\beta_0-0.3586E)}=\frac{\exp(-0.3586E)}{\beta+\exp(-0.3586E)}.
\end{equation}
In equation (\ref{P}), the corresponding values of $\beta = \exp(-\beta_0)$, for each species sensitivity, are listed in Table \ref{tab2}.
The mortality parameters $h$ and $\mu$ are derived from $g(E)$. 
\begin{table}[h]
\caption{Values of $\beta$ to compute $g(E)$ in equation (\ref{P}), corresponding to the considered species sensitivities.}
\label{tab2}\centering
\begin{tabular}{p{3cm}r}
\hline\hline
\\
Sensitivity &   $\beta$ \\ [1ex]
\hline
\textit{extreme}  & $0.0039$ \\ [1ex]
\textit{very high}   & $0.0529$ \\ [1ex]
\textit{high}   & $0.7191$ \\ [1ex]
\textit{above-average} & $9.7747$ \\ [1ex]
\textit{below-average} & $132.9669$ \\ [1ex]
\hline
\end{tabular}
\end{table}
\\
Concretely, Perry et al. compute $\mu$ by numerically integrating  $g(E)$, 
in equation (\ref{P}), between $0$ and $D$, as 
described in
\cite{Perry10}.\\
To estimate the probability of mortality, \emph{h}, for the five hypothetical larvae within the $Bt$-crop, it is necessary to consider that pollen deposition within a maize crop is $2.757$ times that at the edge 
\cite{Perry10}. 
Therefore,
\begin{equation}\label{h}
h = 2.757\;g(0) = 2.757 \frac{1}{1+\beta}.
\end{equation}
The values obtained for $h$, for the considered range of sensitivities, are shown in Table \ref{tab3}. 
\begin{table}[h]
\caption{Within-crop mortality probabilities, $h$, corresponding to the considered species sensitivities. }
\label{tab3}\centering
\begin{tabular}{p{3cm}r}
\hline\hline
\\
Sensitivity &   $h$ \\ [1ex]
\hline
\textit{extreme}  & $2.7463$ \\ [1ex]
\textit{very high}   & $2.6185$ \\ [1ex]
\textit{high}   & $1.6037$ \\ [1ex]
\textit{above-average} & $0.0928$ \\ [1ex]
\textit{below-average} & $0.0075$ \\ [1ex]
\hline
\end{tabular}
\end{table}
\\
To conclude the model overview, two further parameters have to be introduced,
i.e., \emph{mitigation parameter} and  \emph{large scale exposure parameter}. We 
describe only the former, since the latter does not affect our discussion.
\\
Perry et al.
\cite{Perry12}
considered
mitigation measures, too. Indeed, the parameter $w$ is the width in metres of the non-$Bt$ maize strips which represents the simulated mitigation measures which are assumed to be planted around each of the four field edge. When there is mitigation ($w>0$), similar calculations to those just described are used. However, mortality calculated for larvae in the margin must use an appropriate value of $E$, computed to allow for the fact that the $Bt$-maize is a distance $w$ metres further away. The same reasoning is made for the mortality of larvae within the non-$Bt$-maize.   
\subsection{Computation of the non-Bt maize strips width}
\begin{figure}
\begin{center}
  \scalebox{2.0}{\includegraphics{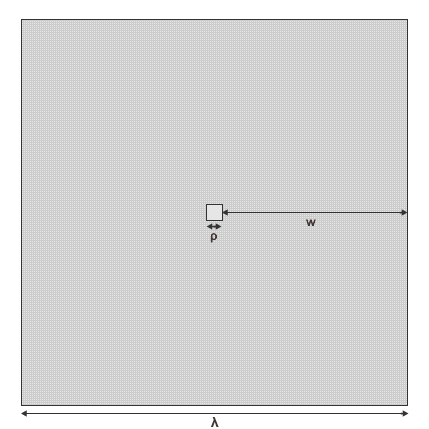}}
 \end{center}
\caption{Crop Field. The parameters $\lambda$, $\rho$, $w$ stand for the size of squared crop field, 
the size of $Bt$-maize cultivated field, the width of non-$Bt$ strips, respectively.  }
\label{circleplot}
\end{figure}

In this Section, we propose a procedure that fixes automatically the width of non  Bt-maize strips.
To this purpose, 
it is necessary to consider that the Cry1F Bt-maize exposure model is  based on the 
Cry1Ab Bt-maize one, previously  described in Section \ref{Cry1ab}.
As a consequence of that, the model inherits from  Cry1Ab Bt-maize exposure model, 
the proportion of population that suffers mortality $P$. 
Therefore,  the analysis regarding
the computation of the exact value of $P$, described in Section \ref{Cry1ab}, 
is equally valid for the Cry1F Bt-maize exposure model, too. This implies that  $P$, given by the equation (\ref{eqfin2}), 
asymptotically tends to $\frac{1}{D}$ where $D$ is the distance from the crop edge.
That being said, we propose a method to assess automatically, starting from the expression
of the equation (\ref{eqfin2}), the width $w$ of the non-Bt maize strips. We propose 
to use the 3$\sigma$ rule \cite{Ross09}, borrowed by the statistical and probability theory,
to derive a fixed value for  the width $w$ . Let be $P(0)$, see equation (\ref{p0}), the proportion of population that suffers mortality 
on the crop edge,  the width $w$ is given by the value of $D$ such that:
\begin{equation}
\frac{P(D)}{P(0)} \le (1- \eta),
\end{equation}
where $\eta$ is the probability that a zero-mean normal distribution with standard deviation $\sigma$ assume values in
$[-3\sigma,3\sigma]$, i.e. $\sim0.9973$.   Since ${P(D)}  \approx O(\frac{1}{D})$, for large $D$, we fix:
\begin{equation}\label{rule}
w  = \{ D :  \frac{1}{D} \approx  (1- \eta) \}
\end{equation}
Hence $D$ can be fixed to $1/(1-0.9973)$, i.e. $\sim 370$m. 
If we use this $D$ value to fix  the width $w$ of the non-Bt maize strips we have the consequence that the Bt-maize field, that we assume for simplicity squared, must have a size larger than twice the value of $D$, i.e., $740$m. For instance,   we consider the example described in \cite{Perry12} with the unique difference that the size of the field is $774$m (the double of the one considered by Perry\footnote{the original size of the field considered by Perry et al., $387$m, cannot be considered since it is lower than $740$m.}). It's easy to show that the area of the field that can be devoted to the Bt-maize cultivation is a square of size $34$m (see Figure \ref{circleplot}), whereas the rest of the field cannot be used since it should be destined to mitigation field.  
The example shows that the adoption of mitigation measures implies strong constraints. Firstly, field size should be larger than $740m$;
secondly,  only a limited field area can be devoted to the $Bt$-maize cultivation.
We conclude the analysis showing how  the values proposed by Perry et al. can be viewed in our
statistical and probabilistic   framework. In particular if it considers the most severe value 
proposed by Perry et al., namely $24m$, after  the rule (\ref{rule}) application it obtains for $\eta$ a value 
of $ \sim0.0416$ that corresponds to apply approximatively a 2$\sigma$ rule 
($2.03 \sigma$ \cite{Korn00} for the precision) . This choice cannot be absolutely considered to fulfill the \emph{worst case scenario}, recommended in the Directive 2001/18/EC  of the European Community \cite{Europ01}.
\subsection{Derivation of a lower bound for species sensitivity}
Now we examine the \textit{within crop mortality} parameter, $h$, 
and its assumed values listed in Table \ref{tab3} of the section \ref{Cry1F}. 
Three out of five $h$ values,  corresponding to the \textit{extreme}, \textit{very high} and \textit{high} sensitivities, violate the definition of probability. 
Recall that \textit{the probability alway takes on values in  between zero and one} \cite{Ross09}, while the above mentioned $h$ values are \textit{greater than one}. 
The joint assumptions that pollen deposition within a maize crop is $2.757$ times that at the edge, and the considered range for 
the species sensitivities do not appear to fit the theoretical soundness of the proposed model. 
The right side of equation (\ref{h}) should be $\le 1$, hence 
equation (\ref{h}) \emph{defines a probability if and only if $\beta\geq1.757$}. 
Therefore, more restrictive assumptions on the sensitivity should be made and/or further argumentation should be given concerning with the assumption that pollen deposition within a maize crop is $2.757$ times that at the edge. 
Obviously, the incorrect values of within-crop mortality $h$, 
for \textit{extreme}, \textit{very high} and \textit{high} sensitivities are propagated 
through the model affecting the correctness of  the proportion of population that suffers mortality $P$ computed 
in equation (\ref{eqfin}) and, consequently, of the entire model. 
However, assuming the above mentioned Perry et al.'s issues are reasoneable, a lower bound for the values of sensitivity $m$
can be added to the Perry et al's model in order to get a correct probability value for the parameter $h$. 
\\
We pass to derive the lower bound.
By applying the definition of probability, the following relations must hold:
\begin{displaymath}
h \leq 1 \Leftrightarrow \frac{2.757}{1+\beta} \leq 1 \Rightarrow \beta \geq 1.757,  
\end{displaymath}   
and, since $\beta=\exp(-\beta_0)$, we get 
\begin{displaymath}
\exp(-\beta_0) \geq 1.757  \Rightarrow \beta_0 \leq -\ln 1.757 \Rightarrow \beta_0 \leq -0.5636.
\end{displaymath}
It comes, from equation (\ref{logit}), that  $\beta_0 = \alpha+5.8017$ and therefore, $\alpha \leq -6.3653$. 
Now, from equation (\ref{logitPd}) we  get:
\begin{equation}
logit(P)-2.473\; \log_{10}d \le -6.353
\end{equation}
and setting $logit(P)=0$, it follows:  
\begin{eqnarray}
-2.473\; \log_{10}d  & \leq  &-6.3653 \nonumber \\ 
 \log_{10}d & \geq & 2.5739 \nonumber \\ 
d & \geq & 10^{2.5739} \sim 374.89.
\end{eqnarray}
Therefore, since at $LC50$, $m=d$ (see supplementary material $S1$ and $S3$ in \cite{Perry12}),
the lower bound $m\geq374.89$ must hold in order to $h$ to be a probability.
\section{Conclusion}\label{concl}
In this work, some mathematical models for the estimation of the effects of Cry1Ab and Cry1F Bt-maize exposure on 
non-target Lepidoptera  have been discussed, deriving  novel results. Firstly, it is obtained the  exact formula for  the proportion of the population that suffers mortality exposed to Cry1Ab pollen, studying its  dependence  on the margin from the $Bt$-crop edge.
Besides,  regarding Cry1F pollen effects, a  procedure was proposed to fix automatically  the width of the non $Bt$-stripes used as mitigation measures. It was also shown that the adoption of mitigation measures, that take into account the worst-case scenario
recommended in the directive  2001/18/EC by European Community, implies strong constraints
on the Bt-maize cultivation. Firstly, field size should be larger than $740m$; secondly,  only a limited field area can be devoted to the $Bt$-maize cultivation.
Furthermore, on the basis of  probabilistic considerations,  it has been derived a lower bound, on the species 
sensitivity of  Lepidoptera.
\\
We hope that the derived novel results  about mathematical models for the estimation of the effects of Cry1Ab and Cry1F Bt-maize exposure on  non-target Lepidoptera  can contribute  to the debate in the scientific community about the GMP effects.
%

%
%
\end{document}